\documentclass[1p,number]{elsarticle}

\usepackage{amsmath}
\usepackage{amsfonts}
\usepackage{amssymb}

\usepackage{graphicx}
\usepackage{caption}
\usepackage{subcaption}

\newcommand{\norm}{\mathcal{N}}
\newcommand{\e}{\mathrm{e}}
\newcommand{\order}{\mathcal{O}}
\newcommand{\Dt}{\Delta t}

\journal{Journal of Computational Physics}

\begin{document}

\begin{frontmatter}

\title{Simulating DNLS models}

\author{Mario Mulansky}

\address{University of Potsdam, Department of Physics and Astronomy}
\address{Louisiana State University, Center for Computation and Technology ({\tt mmulansky@cct.lsu.edu})}

\begin{abstract}
We present different techniques to numerically solve the equations of motion for the widely studied Discrete Nonlinear Schr\"odinger equation (DNLS).
Being a Hamiltonian system, the DNLS requires symplectic routines for an efficient numerical treatment.
Here, we introduce different such schemes in detail and compare their performance and accuracy by extensive numerical simulations.
\end{abstract}

\begin{keyword} 
Symplectic Integration; Performance; DNLS; Hamiltonian System
\end{keyword}

\end{frontmatter}


\section{Introduction}

The Discrete Nonlinear Schr\"odinger Equation (DNLS) has been known to physicists, biologists, chemists and mathematicians for more than 30 years now.
It describes a simple model of coupled unharmonic oscillators defined by the complex valued phase/amplitude $\psi_n\in\mathbb{C}$ at lattice site $n$.
The dynamics of the oscillators is governed by the DNLS that writes in one spatial dimension:
\begin{equation} \label{eqn:DANSE}
 i \frac\partial{\partial t} \psi_n = V_n \psi_n + \psi_{n+1} + \psi_{n-1} + \beta |\psi_n|^2 \psi_n,
\end{equation}
where $i$ is the imaginary unit, $V_n\in\mathbb{R}$ is the local potential and $\beta\in\mathbb{R}$ the nonlinear strength.
These equations of motion can be derived from the following Hamilton function:
\begin{equation} \label{eqn:Hamilton}
 H = \sum\limits_{n} V_n|\psi_n|^2 + \psi_{n+1}^* \psi_n + \psi_{n+1} \psi_n^* + \frac\beta2 |\psi_n|^4.
\end{equation}
The Hamiltonian character immediately gives an integral of motion which is usually called energy and denoted as $E:=H=\text{const}$.
However, the DNLS possesses another conserved quantity, the norm $\norm:= \sum_n |\psi_n|^2$, usually set to unity $\norm=1$.
The existence of two conserved quantities makes this equation particularly challenging for numerical approaches, as will be explained later.

This model first appeared as a description of polarons by Holstein~\cite{Holstein_59}.
Later, the DNLS was used by Davydov in his studies of protein dynamics~\cite{Davydov_73,Scott_82} as well as in the context of local modes in small molecules~\cite{Scott_Eilbeck_86}.
In recent years, however, two new applications of the DNLS gained increasing attention: Bose-Einstein-Condensates (BEC) and coupled optical wave guides.

In the context of Bose-Einstein condensates, the DNLS describes the mean field approximation of a weakly interacting Bose gas at zero temperature as shown e.g.\ in~\cite{Mishmash_Carr_09}.
The predictions of this description have been, to some extent, experimentally verified for condensates in optical traps and a periodic potential, e.g.~\cite{Anderson_Kasevich_98,Cataliotti_etal_01}.
A very interesting phenomenon that can be studied in BECs is Anderson localization, which denotes the ``absence of diffusion''~\cite{Anderson_58} in disordered systems.
Several experimental realizations of this effect have been reported, e.g.~\cite{Schwartz_etal_07,Sanchez-Palencia_etal_07,Lahini_etal_08,Billy_etal_08,Roati_etal_08}.
The corresponding DNLS accompanied by a random local potential is often called the Discrete Anderson Nonlinear Schr\"odinger Equation (DANSE) and has been subject of heavy numerical studies in the past years.

Moreover, the DNLS gives a very accurate description of the propagation of light in optical waveguides with a nonlinear material~\cite{Jensen_80,Christodoulides_Joseph_88}.
As for BECs, optical waveguides were also used to study Anderson localization by imposing a disordered potential.
Especially the fact that the nonlinear strength can be controlled quite precisely in terms of the properties of the nonlinear material makes this system very appealing for studying Anderson localization and nonlinearity experimentally.
This has been done recently by Schwartz et al.\ in~\cite{Schwartz_etal_07}.
A more extensive review on the history and applications of the DNLS can be found in~\cite{Eilbeck_Johansson_03}.

The increasing experimental accessibility of systems described by the DNLS has led to numerous numerical investigations of such models.
Special attention has been given to the case where Anderson localization, induced by disorder, is accompanied by nonlinearity.
In terms of the DNLS this can be modeled by choosing the local potential $V_n$ to be a random variable, typically as independent and identically distributed (iid) from some interval of size $W$: $V_n\in[-W/2,W/2]$.
A large number of numerical investigations studied the slow destruction of Anderson localization by the nonlinear interactions between the localized eigenmodes~\cite{Pikovsky_Shepelyansky_08,Flach_Krimer_Skokos_09,Mulansky_Pikovsky_10,Laptyeva_etal_10,Ivanchenko_Laptyeva_Flach_11,Veksler_Krivolapov_Fishman_09}.
In all these works, subdiffusive spreading of initially localized modes is reported up to computationally accessible times, recently confirmed by an experimental study~\cite{Lucioni_etal_11}.
Additionally, similar spreading laws have been observed in quasi-periodic nonlinear lattices~\cite{Larcher_etal_12} and the nonlinear Stark ladder~\cite{GarciaMata_Shepelyansky_09,Krimer_Khomeriki_Flach_09}.
Recent results on the scaling properties of chaos~\cite{Pikovsky_Fishman_11} and on the scaled spreading in a reduced model~\cite{Mulansky_Pikovsky_11} raised some questions on the asymptotic validity of the observed subdiffusive spreading.
Several attempts to find a theoretical description of the spreading have been proposed, mainly based on an effective noise theory~\cite{Flach_Krimer_Skokos_09}, but a full understanding of the interplay between disorder and nonlinearity is still lacking~\cite{Fishman_Krivolapov_Soffer_12}.
Lately, the theory of chaotic diffusion has been developed~\cite{Mulansky_phd_12} and successfully applied to a different class of systems with disorder and nonlinearity~\cite{Mulansky_Pikovsky_12arxiv,Mulansky_Pikovsky_12}, but can possibly be extended to the DNLS/DANSE models as well.
Here, however, we are not going to further address the very interesting problem of asymptotic spreading, but rather focus on the different algorithmic techniques that are used to obtain those numerical results.

This article is organized as follows: After this introduction, we will review the general properties of symplectic integrators and the idea of operator splitting to construct such methods in section~\ref{sec:symplectic_schemes}.
In section~\ref{sec:numerical_methods} we introduce the different numerical approaches for simulating the DNLS model and in section~\ref{sec:performance} we compare their performance.
Finally, we end with our conclusions in section~\ref{sec:conclusions}.

\section{Symplectic Schemes} \label{sec:symplectic_schemes}

Solving the DNLS model here means to find an approximate solution~$\psi(t)$ of the equations of motion~\eqref{eqn:DANSE} from an initial condition $\psi(t=0)$.
Being a Hamiltonian system, the DNLS has a symplectic flow map and exhibits conservation of energy $E=\text{const}$.
This requires the usage of symplectic routines that preserve the symplectic nature of the system~\cite{Leimkuhler_Reich_05,Hairer_Lubich_Wanner_06}.
Formally, such a solution can be written in terms of the Liouville operator $\e^{tL_H}$, defined via Poisson brackets~\cite{Goldstein_02} and acting on the initial condition $\psi (t) = \e^{tL_H} \psi(0)$.
If the action of this operator is known explicitly, the system is called integrable and the solution $\psi(t)$ can be written analytically.
In most cases, however, this solution can only be approximated by numerical methods.

A very common technique to find a symplectic scheme for a non-integrable system is the so-called operator splitting~\cite{McLachlan_Quispel_02}.
For this, the Hamilton function has to be separable, which means it can be written as $H=A+B$, where the action of the operators $\e^{tL_a}$ and $\e^{tL_B}$ are known explicitly.
Then, the integration of the solution from time $t$ to $t+\Delta t$ for some small time step $\Delta t$ can be approximated by:
\begin{equation} \label{eqn:trivial_splitting}
 \e^{\Delta t L_H} = \e^{\Delta t(L_A+L_B)} = \e^{\Delta t L_A} \e^{\Delta t L_B} + \mathcal{O}(\Delta t^2).
\end{equation}
The splitting scheme above is the most simple and only accurate in first order of $\Delta t$.
In a more general way, the approximation can be written as a product of $j$ operators:
\begin{equation} \label{eqn:general_splitting}
 \e^{\Delta t L_H} = \prod_{i=1}^j \e^{a_i \Delta t L_A} \e^{b_i \Delta t L_B} + \mathcal{O}(\Delta t^{p+1}).
\end{equation} 
The operators $\e^{a_i \Delta t L_A}$ and $\e^{b_i \Delta t L_B}$ represent the exact integrations according to Hamiltonians $A$ and $B$ over times $a_i\Delta t$ and $b_i\Delta t$ respectively.
The parameters $a_i, b_i$ are chosen such that the resulting product is an exact representation of $\e^{\Delta t L_H}$ of order $\Delta t^p$.
Such a splitting scheme effectively computes a trajectory not of the original system with Hamiltonian $H=A+B$, but of a new Hamiltonian ${K=A+B+\mathcal{O}(\Delta t^{p})}$.
Thus, the symplectic property of the dynamics is preserved.
Moreover, the computed trajectory, obtained from many subsequent steps~\eqref{eqn:general_splitting}, conserves a new energy $\tilde E := K = E+\order(\Delta t^p)$, which means that along this trajectory the original energy is not exactly conserved, but only fluctuates around its exact value with a magnitude of $\order(\Delta t^p)$.
With a non-symplectic integration scheme this is not the case, as there the numerical error of the energy increases at every step by an amount $\order(\Delta t^p)$, which accumulates along the trajectory.
Hence, especially for long integration times the symplectic routines are essential as they allow reasonable energy conservation even for large time steps $\Delta t$ which makes them much more efficient than non-symplectic methods.
Thus, symplectic integrators have become standard for Hamiltonian systems and several splitting routines of different orders have been developed in recent years, see e.g.~\cite{Yoshida_90,Candy_Rozmus_91,McLachlan_95,Chin_97,Channell_Scovel_99,Laskar_Robutel_01}.

In the following, we will use two second order ($p=2$) symmetric splittings: the SBAB$_1$ and SBAB$_2$ scheme~\cite{Laskar_Robutel_01}.
Symmetric splittings are especially interesting as they naturally lead to self-adjoint algorithms.
The time evolution according to those splittings for a separated Hamilton function $H=A+B$ are defined as follows:
\begin{align}
 \text{SBAB}_1: \e^{\Delta tL_H} &= \e^{\Delta t/2 L_B} \e^{\Delta t L_A} \e^{\Delta t/2 L_B} + \mathcal{O}(\Delta t^3) \label{eqn:SBAB1} \\
   K &= A + B + \order(\Dt^2\cdot|B|) \nonumber \\
 \text{SBAB}_2: \e^{\Dt L_H} &= \e^{b_1\Delta t L_B} \e^{a_1\Delta t L_A} \e^{b_2\Delta t L_B} \e^{a_1\Delta t L_A} \e^{b_1\Delta t L_B} + \mathcal{O}(\Delta t^3) \label{eqn:SBAB2} \\
   K &= A + B + \order(\Dt^2\cdot|B|^2), \nonumber
\end{align}
with $a_1 = 1/2$ , $b_1 = 1/6$ and $b_2 = 2/3$ in the last line and $K$ denoting the formal Hamiltonian integrated exactly by the given scheme.
The symmetric composition ensures that these schemes are identical to their inverse with negative timestep: $\text{SBAB}_{1/2}(\Dt) = (\text{SBAB}_{1/2}(-\Dt))^{-1}$, a property denoted as self-adjoint.

\section{Numerical Methods} \label{sec:numerical_methods}

Here, we will present different algorithms to numerically calculate approximate trajectories of the DNLS equation.
At first we focus on methods based on the two-part operator splitting described above, all of which have a different approach to solve the non-local term in~\eqref{eqn:DANSE}.
The first method uses Fourier transform for the non-local part and has been applied to the DNLS in~\cite{Skokos_etal_09,Flach_Krimer_Skokos_09,Laptyeva_etal_10,Ivanchenko_Laptyeva_Flach_11}, it will be abbreviated as ``FT'' here.
The second algorithm employs an implicit Crank-Nicolson scheme and has been used in~\cite{Mulansky_dipl_09,Mulansky_Pikovsky_10} to integrate the DNLS, it will be referred to as ``CN'' throughout this text.
The third method is the so-called ``PQ'' scheme introduced in~\cite{Bodyfelt_etal_11}, where the non-local part is again separated into two integrable operators.
Finally, a new multi-symplectic scheme is introduced based on the ideas of Bridges and Reich~\cite{Bridges_Reich_01} that was already applied to a DNLS without local potential~\cite{Wang_Li_Song_08} and will be called Euler-Box ``EB'' scheme here.

\subsection{Fourier Method}

The obvious way to treat the coupling term in the DNLS is to use a spectral method.
This involves forward and backward Fourier transforms at every time step and is thus called ``FT'' in this text.
It has already been applied to the DNLS as described in~\cite{Skokos_etal_09}.
For this scheme, the Hamilton function~\eqref{eqn:Hamilton} is split into two parts $H=A_{FT}+B_{FT}$ with:
\begin{equation} \label{eqn:ft_H_splitting}
\begin{aligned}
 A_{FT} &= \sum\limits_n \psi_{n-1}\psi_n^* + \psi_n\psi_{n-1}^* \\
 B_{FT} &= \sum\limits_n V_n |\psi_n|^2 + \frac\beta2 |\psi_n|^4.
\end{aligned}
\end{equation}
Here, $A$ contains only the coupling and $B$ consists of the linear and nonlinear local potential.
This implies that the action of the time evolution operator $\e^{\Dt L_{A_{FT}}}$ becomes local when applied to the Fourier transform of the state~$\tilde \psi_q$.
Specifically, $\e^{\Dt L_{A_{FT}}}$ acts as follow~\cite{Skokos_etal_09}:
\begin{equation}
 \e^{\Dt L_{A_{FT}}}: \begin{cases}\begin{aligned}
                \tilde\psi_q &= \sum_{n=1}^N \psi_n \e^{2\pi i q (n-1)/N}\\
                \tilde\psi_q' &= \e^{-2i\Dt\cos( 2\pi(q-1)/N )} \tilde\psi_q \\
		\psi_n' &= \frac1N \sum_{n=1}^N \tilde\psi_q' \e^{-2\pi i n (q-1)/N},
               \end{aligned}\end{cases}
\end{equation}
where the first and the last operation are forward- and backward Fourier transforms of the state $\psi_n$ and the second step is the time evolution in Fourier space.
The action of $\e^{\Dt L_{B_{FT}}}$ can be written explicitly as it is fully local in real space:
\begin{equation}
 \e^{\Dt L_{B_{FT}}}: \psi_n' = \e^{-i\Dt(V_n + \beta|\psi_n|^2)} \psi_n.
\end{equation}

\begin{figure}[t]
  \centering
    \includegraphics[width=0.6\textwidth]{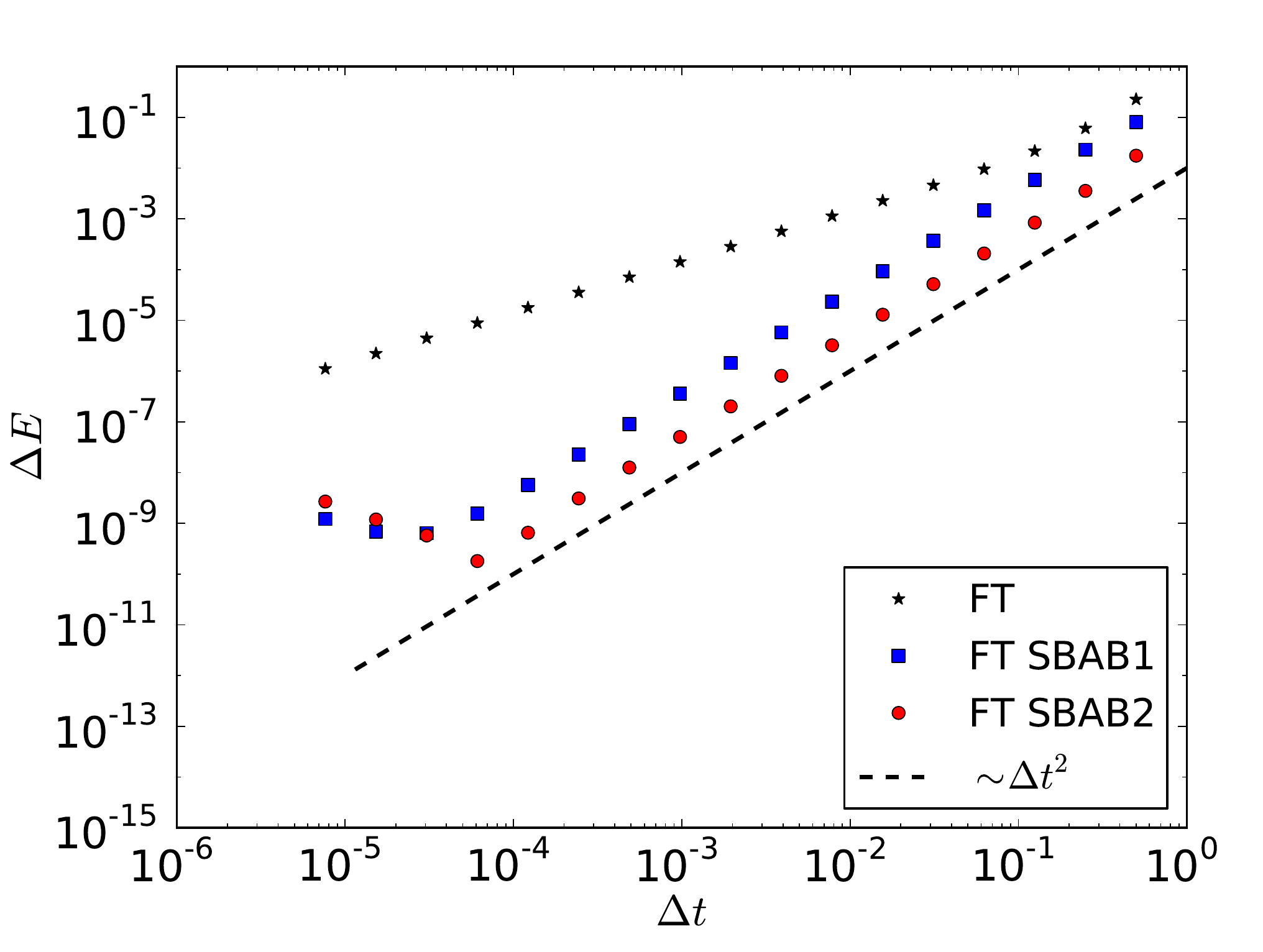} \hfill
  \caption{Behavior of the energy error for the FT scheme and different splitting methods. The simulation was done for a DNLS with $N=128$ lattice sites, $\beta=1$ and a random potential $V_n\in[-2,2]$.
  $\Delta E$ is the averaged energy error along the trajectory~\eqref{eqn:dE} up to a total integration time~$T=100$.}
   \label{fig:ft}
\end{figure}

Having found two explicit representations of the two parts of the time evolution one can now construct integrators by consecutively applying $\e^{\Dt L_{A_{FT}}}$ and $\e^{\Dt L_{B_{FT}}}$.
The simplest form is given by the first order approximation $\e^{\Dt L_H} = \e^{\Dt L_{A_{FT}}} \e^{\Dt L_{B_{FT}}} +\order(\Dt^2)$.
As described above, symmetric second order schemes can be obtained by using the SBAB$_1$ or SBAB$_2$ splitting~\eqref{eqn:SBAB1}, \eqref{eqn:SBAB2}.
A numerical study on the behavior of the energy error for the simple first order and the two second order splitting schemes of the FT method is shown in Figure~\ref{fig:ft}.
We compared the three splitting methods by integrating the DNLS system~\eqref{eqn:DANSE} with a random potential chosen independent and identically distributed $V_n \in [ -2,2]$, a nonlinear strength of $\beta=1$ and $N=128$ lattice sites -- the DNLS with a random potential is usually referred to as DANSE model.
As initial condition we chose a Gaussian in the lattice center with width~$\sigma=10$, ensuring $\norm = \sum |\psi_n|^2 = 1$.
We performed the numerical time evolution using each of the above schemes up to the time $T=100$ and repeated for decreasing step sizes $\Dt = 1.0 \dots 10^{-6}$.
As quantification for the accuracy we calculate the mean square energy error $\Delta E$ and norm $\Delta\norm$ along the trajectory:
\begin{equation} \label{eqn:dE}
 \Delta E = \sqrt{\langle (E_m-E_0)^2} \rangle = \sqrt{\frac1M \sum_m (E_m-E_0)^2}, \qquad 
 \Delta\norm = \sqrt{\langle (\norm_m-\norm_0)^2} \rangle 
\end{equation} 
where the index $m$ denotes the time and $E_0$ is the exact energy and $\norm_0=1$ the exact norm, fixed to unity in these simulations.
Figure~\ref{fig:ft} shows the behavior of the energy error for the different splittings for decreasing stepsize.
One clearly observes the second order behavior of the SBAB$_1$ and SBAB$_2$, compared to the first order trivial splitting~\eqref{eqn:trivial_splitting}.
Additionally, the SBAB$_2$ splitting shows a significantly smaller energy error than the simpler ABC and SBAB$_1$ schemes.
For stepsizes smaller than ${\Dt \sim 10^{-4}}$ the finite double precision starts limiting the accuracy of the computation, hence smaller stepsizes $\Dt < 10^{-4}$ are not advisable for the SBAB$_{1/2}$ FT schemes.
Note, that the FT based splitting schemes exhibit numerically exact norm conservation so no analysis of $\Delta \norm$ is required.

\subsection{Crank-Nicolson (CN)}
As the Schr\"odinger equation is of such great importance, many numerical methods have been developed to find trajectories in this system.
One of the most important approaches is the ``Crank-Nicolson'' scheme~\cite{Crank_Nicolson_96}, a second order symplectic scheme to treat the usual (linear) Schr\"odinger equation, also used for solving other partial differential equations like the wave equation~\cite{NumRecCpp}.
To use this scheme, we will split the Hamilton function~\eqref{eqn:Hamilton} into a linear and a nonlinear part and then use the ideas of Crank and Nicolson to solve the linear part.
Hence, we write $H=A_\text{CN}+B_\text{CN}$ with:
\begin{equation}  \label{eqn:cn_H_splitting}
\begin{aligned}
 A_\text{CN} &= \sum\limits_n V_n |\psi_n|^2 + \psi_{n-1}\psi_n^* + \psi_n\psi_{n-1}^* \\
 B_\text{CN} &= \sum\limits_n \frac\beta2 |\psi_n|^4.
\end{aligned}
\end{equation}

The Liouville operators for the two individual terms are then defined by their actions on $\psi$ as:
\begin{align}
 \dot\psi_n = (L_{A_\text{CN}} \psi)_n & = -i( \psi_{n+1} + \psi_{n-1} + V_n\psi_n) \label{eqn:linearPart}\\
 \dot\psi_n = (L_{B_\text{CN}} \psi)_n & = -i \beta |\psi_n|^2\psi_n. \label{eqn:nonlinearPart}
\end{align}
The time evolution operator $\e^{\Dt L_{B_\text{CN}}}$ is nonlinear, but can then be written explicitly due to its local character:
\begin{equation}
 \e^{\Dt L_{B_\text{CN}}}: \psi_n' = \e^{-i\Dt\beta|\psi_n|^2} \psi_n,
\end{equation}
where $\psi'$ denotes the time evolution of one time step $\Dt$ according to the Hamilton function $B$ only.

Unfortunately, $\e^{\Dt L_{A_\text{CN}}}$ can only be written explicitly for a constant or periodic potential.
As we want to treat the generic case here, we will use the Crank-Nicolson scheme, a second order, norm preserving approximative method.
Applying it to the linear part $L_A$~\eqref{eqn:linearPart} (we skip the index CN her for simplicity), the exponential is approximated by Caley's formula: ${\mathrm{e}^{\Delta tL_{A}} \approx \frac{1 + L_A\Delta t/2}{1 - L_A\Delta t/2}}$, which leads to an implicit equation for the evolved state $\psi'$:
\begin{equation} \label{eqn:crank-nicolson}
 \e^{\Dt L_A}: (1 - L_A\Delta t/2) \psi' = (1 + L_A\Delta t/2) \psi.
\end{equation}
From \eqref{eqn:linearPart} can be seen that $L_A$ is a band matrix with $-iV_n$ on the diagonal and $-i$ on the upper/lower sub-diagonal, this means \eqref{eqn:crank-nicolson} is a set of linear equations which can be solved by a Gau\ss-algorithm.
Moreover, due to the band structure of $L_A$ the computational requirement is only linear in the system size.

\begin{figure}[t]
  \centering
    \includegraphics[width=0.6\textwidth]{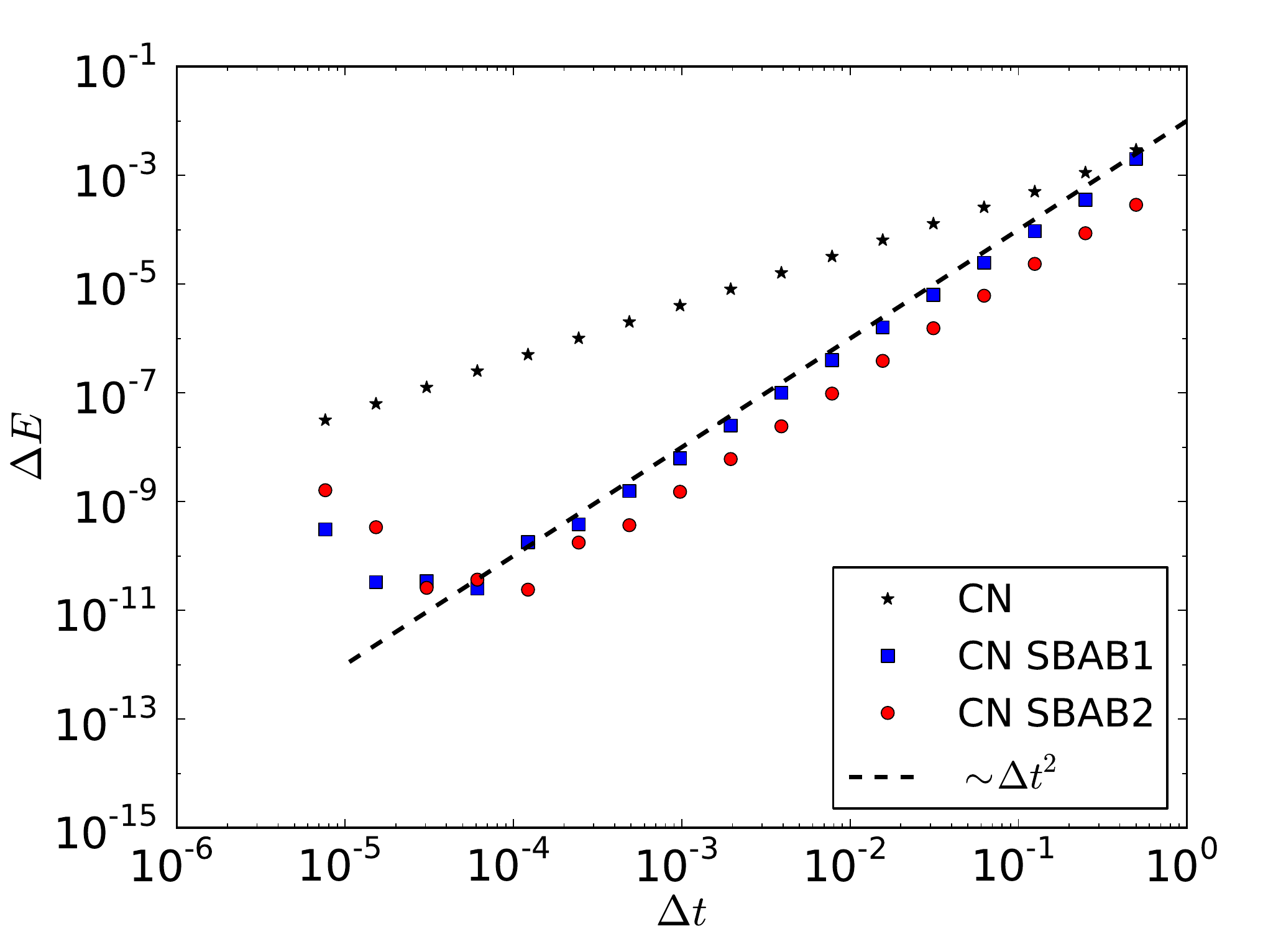} \hfill
  \caption{Behavior of the energy error for the CN scheme and different splitting methods. The simulation setup is as described in Figure~\ref{fig:ft}. The dashed line shows exemplarily the second order behavior and is located at the same position as in Figure~\ref{fig:ft} for comparison.}
  \label{fig:cn}
\end{figure}

Using the symplectic integrators for the two parts $A_\text{CN}$ and $B_\text{CN}$ one can now construct a symplectic scheme as described in section~\ref{sec:symplectic_schemes}.
Namely, we have implemented the usual Crank-Nicolson splitting scheme of order $p=1$: $\text{CN}$ after~\eqref{eqn:trivial_splitting}, the second order $\text{CN}_\text{SBAB1}$ scheme following~\eqref{eqn:SBAB1} and the $\text{CN}_\text{SBAB2}$ scheme from~\eqref{eqn:SBAB2}.
Remember that all these schemes are symplectic and norm preserving.
We also note that for weak nonlinearities the last method $\text{CN}_\text{SBAB2}$ represents a considerable improvement over $\text{CN}_\text{SBAB1}$, as the error gets significantly smaller when $|B_\text{CN}| \ll 1$.
In a typical spreading setup as chosen in the numerical examples of this article, the total error in the energy reduces by about one order of magnitude, as seen in Figure~\ref{fig:cn}.

The results are presented in Figure~\ref{fig:cn} in terms of the energy error~$\Delta E$, as the norm again is preserved exactly.
The plots illustrate the respective $p=1,2$ behavior of the methods, as well as the improvement of the $\text{CN}_\text{SBAB2}$ over the more simple $\text{CN}_\text{SBAB1}$.
For both methods we found an optimal stepsize~$\Dt \approx 10^{-4}$ that leads to minimal energy errors of $\Delta E\approx 10^{-12}$.
For stepsizes below these values, the finite double precision again leads to a linear increase of the error.
Furthermore, the error of the CN$_\text{SBAB2}$ is clearly smaller than for the FT$_\text{SBAB2}$ shown in Figure~\ref{fig:ft}, as easily seen from comparing with the dashed black lines that are plotted at the same position in both graphs.

\subsection{PQ-Splitting}
While in the methods above Fourier transforms or the Crank-Nicolson scheme were used as an integrator of the linear coupling term in $H$, Bodyfelt et al.\ presented a new idea to treat the nonlocal term -- the PQ scheme~\cite{Bodyfelt_etal_11}.
They proposed to apply another operator splitting of the linear part such that the final evolution operator consists of three separate mappings.

\begin{figure}[t]
  \centering
    \includegraphics[width=0.6\textwidth]{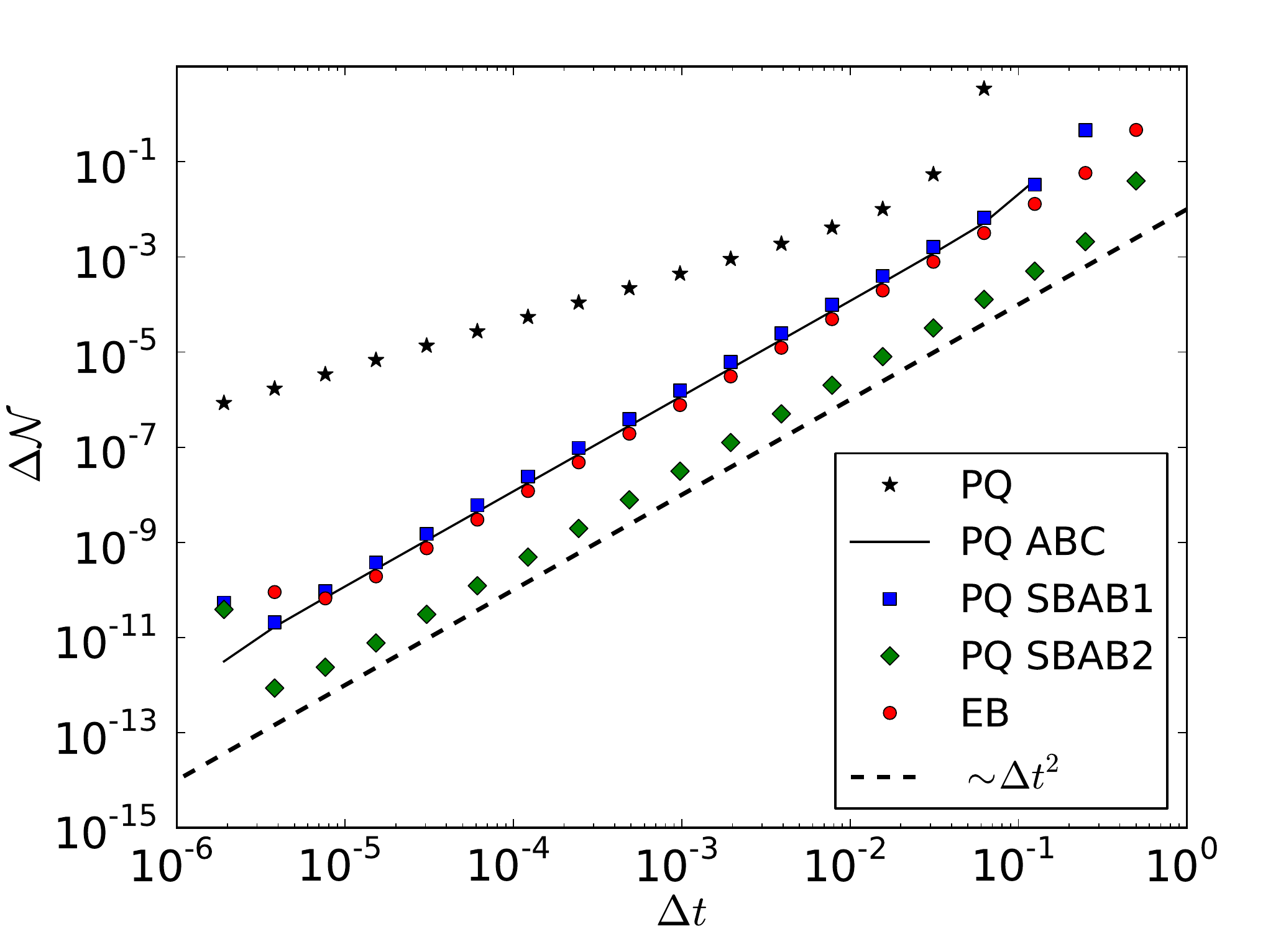} \hfill
  \caption{Behavior of the norm error for different splitting methods for the PQ scheme and the Euler-Box method. The simulation setup is as described in Figure~\ref{fig:ft}. The dashed line shows exemplarily the second order behavior and is located at the same position as in Figures~\ref{fig:ft} and \ref{fig:cn} for comparison.}
  \label{fig:pq}
\end{figure}

To derive this method, we first have to divide the state into real and imaginary part: $\psi_n = a_n + ib_n$.
The Hamilton function \eqref{eqn:Hamilton} is then:
\begin{equation}
 H = \sum_n V_n(a_n^2+b_n^2) + 2(a_n a_{n+1} + b_n b_{n+1}) + \frac\beta2 (a_n^2+b_n^2)^2.
\end{equation} 
This can then be split into the three integrable parts:
\begin{align} \label{eqn:pq_splitting}
 B =& \sum_n V_n(a_n^2+b_n^2) + \frac\beta2 (a_n^2+b_n^2)^2 \nonumber \\
 Q =& \sum_n 2 a_n a_{n+1} \\
 P =& \sum_n 2 b_n b_{n+1}. \nonumber 
\end{align}
As above, the time evolution operator can now be approximated: 
\begin{equation} \label{eqn:hpq_splitting}
 \text{PQ}: \mathrm{e}^{\Dt H} = \e^{\Dt L_B} \e^{\Dt L_Q} \e^{\Dt L_P} + \order(\Dt^2),
\end{equation} 
where with defining $\alpha_n = 2V_n + \beta ( a_n^2 + b_n^2 )$ the individual steps are given as:
\begin{align}
 \e^{\Dt L_B} :& 
  \begin{cases}
    a_n' = a_n\cos(\alpha_n \Delta t) + b_n\sin(\alpha_n \Delta t) \\
    b_n' = b_n\cos(\alpha_n \Delta t) - a_n\sin(\alpha_n \Delta t)
  \end{cases} \label{eqn:PQ_L_B} \\
 \e^{\Dt L_Q} :& 
  \begin{cases}
   a_n' = a_n \\
   b_n' = b_n - 2\Delta t ( a_{n-1} + a_{n+1} )
  \end{cases} \label{eqn:PQ_L_Q} \\
 \e^{\Dt L_P} :& 
  \begin{cases}
   a_n' = a_n + 2\Delta t ( b_{n-1} + b_{n+1} )\\
   b_n' = b_n,
  \end{cases} \label{eqn:PQ_L_P}
\end{align}
This concatenation of three solvable evolution operators defines a first order symplectic scheme which is an integrator of the formal Hamiltonian ${K = B+P+Q + \order(\Dt)}$.
However, as pointed out in section~\ref{sec:symplectic_schemes}, the error of the splitting can be improved by using better approximations than \eqref{eqn:hpq_splitting}.
The simplest symmetric second order splitting is given by:
\begin{align} \label{eqn:pq_abc}
 \text{PQ}_\text{ABC}: \e^{\Dt L_H} &= \e^{\frac\Dt2 L_B} \e^{\frac\Dt2 L_P} \e^{\Dt L_Q} \e^{\frac\Dt2 L_P} \e^{\frac\Dt2 L_B} + \mathcal{O}(\Delta t^3) \\
   K &= B + P + Q + \order(\Dt^2 \cdot |B||P||Q|). \nonumber
\end{align}
Surprisingly, this obvious three-operator concatenation has been discussed only very recently by Skokos et al.~\cite{Skokos_etal_13} where it also was applied to the DNLS.
It was called ``ABC'' scheme there, so we denote it as PQ$_\text{ABC}$ here.
Besides that, one can also use the two-operator splitting presented above and apply that two times.
This idea was used in~\cite{Bodyfelt_etal_11} where the Hamiltonian $H$ is split in two steps as follows: first $H = A+B$ followed by $A=P+Q$.
Applying the Leapfrog method $\text{SBAB}_1$ to the two splittings we get the following scheme:

\begin{align} \label{eqn:pq_sbab1}
 \text{PQ}_\text{SBAB1}: \e^{\Dt L_H} &= \e^{\frac\Dt4 L_B} \e^{\frac\Dt2 L_P} \e^{\frac\Dt4 L_B} \e^{\Dt L_Q} \e^{\frac\Dt4 L_B} \e^{\frac\Dt2 L_P} \e^{\frac\Dt4 L_B} + \mathcal{O}(\Delta t^3) \\
   K &= B + P + Q + \order(\Dt^2 \cdot |B||P||Q|). \nonumber
\end{align}
Applying the $\mathrm{SBAB}_2$ method individually to the two splittings also results in a second order symmetric scheme given by the following 13 successive simple mappings:
\begin{align}
 \text{PQ}_\text{SBAB2}: \mathrm{e}^{\Delta t H} =&\; \mathrm{e}^{d_1\Delta tL_B} \mathrm{e}^{d_1c_2\Delta tL_P} \mathrm{e}^{c_2^2\Delta tL_Q} \mathrm{e}^{d_2c_2\Delta tL_P} \mathrm{e}^{c_2^2\Delta tL_Q} \nonumber \\
  & \times \mathrm{e}^{d_1c_2\Delta tL_P} \mathrm{e}^{d_2\Delta tL_B} \mathrm{e}^{d_1c_2\Delta tL_P} \mathrm{e}^{c_2^2\Delta tL_Q} \\
  &  \times \mathrm{e}^{d_2c_2\Delta tL_P} \mathrm{e}^{c_2^2\Delta tL_Q} \mathrm{e}^{d_1c_2\Delta tL_P} \mathrm{e}^{d_1\Delta tL_B} + \mathcal{O}(\Delta t^3) \nonumber \\ 
  K &= B + P + Q + \order(\Dt^2 \cdot |B|^2|P|^2|Q|) \nonumber
\end{align} 
with $d_1=1/6$, $d_2=2/3$ and $c_2=1/2$ \cite{Laskar_Robutel_01}.
This is again symplectic and has order $p=2$.
For small $B$ this would lead to reduction of the error.
However, for the splitting~\eqref{eqn:pq_splitting}, the norms of the operators $B$ , $P$ , $Q$ are all $\sim 1$.
Hence, it is not obvious that this scheme leads to a better error behavior, but a numerical investigation shows indeed that the energy error $\Delta E$ decreases by more than one order of magnitude when using the $\text{SBAB}_2$ method compared to $\text{SBAB}_1$.
This is shown in Figure~\ref{fig:pq}, where we again integrated the DNLS~\eqref{eqn:DANSE} as above with $N=128$ sites, an idd.\ random potential $V_n = -2\dots2$ and $\beta=1$ up to the time $T=100$ using decreasing time steps $\Dt = 0.5 \dots 10^{-6}$.
As initial conditions we again used a Gaussian at the center with width~$\sigma=10$.
Because the norm is not exactly conserved by the PQ scheme, we will here use $\Delta\norm$ as quantification for the accuracy of the method for the sake of variety.
One finds that the $\text{PQ}_\text{ABC}$ and the $\text{PQ}_\text{SBAB1}$ produce very similar errors.
The Euler-Box scheme, presented in the next chapter, exhibits a slightly smaller error, but by far the best splitting is the $\text{PQ}_\text{SBAB2}$.
Comparing Figures~\ref{fig:cn} and \ref{fig:pq} one finds that the CN based schemes still produce the smallest errors.
However, the PQ schemes are computationally much simpler which allows smaller time steps $\Dt$ at the same CPU time compared to the CN methods.
Hence, a detailed performance study is required to identify which method is more efficient.
This will be done in section~\ref{sec:performance}.

\subsection{Multi-symplectic Euler-Box Scheme} \label{sec:euler_box}
The above schemes relied on the splitting of the Hamilton function into integrable parts.
In this section, we will present a different approach that is based on a multi-symplectic formulation of the equations.
Multi-symplectic schemes were introduced by Bridges and Reich~\cite{Bridges_Reich_01} for Hamiltonian PDEs and they are known to locally conserve symplectivity.
As the DNLS system can be thought of as a discretized version of a continuous nonlinear Schr\"odinger equation, the application of the multi-symplectic theory is reasonable~\cite{Islas_Karpeev_Schober_01}.
Actually, this has been done already by Wang et\ al.~\cite{Wang_Li_Song_08} where a multi-symplectic scheme for the DNLS without local potential ($V_n=0$) was developed.
Here, we will adopt this scheme closely following the calculations in~\cite{Wang_Li_Song_08} for the general DNLS~\eqref{eqn:DANSE}.
As result, we will obtain a locally implicit, self-adjoint multi-symplectic scheme of second order.

To derive this scheme we have to start at the continuous wave equation for the complex valued, space- and time-dependent wave function $\psi(x,t)$:
\begin{equation} \label{eqn:danse_continuous}
 i\psi_t = \psi_{xx} + 2\psi + V(x)\psi + \beta |\psi|^2 \psi.
\end{equation}
Note, that from a straight forward spatial discretization with grid size $\Delta x = 1$ one immediately obtains~\eqref{eqn:DANSE}.
By setting $\psi(x,t) = a + ib$ and $v=a_x$, $w=b_x$ this can be formulated as:
\begin{equation}
 M z_t + K z_x = \nabla_z S(z),
\end{equation}
with 
\begin{equation}
  S = \frac{2+V}2(a^2+b^2) + \frac12(v^2+w^2) + \frac\beta4( a^2 + b^2 )^2
\end{equation}
and $M$ and $K$ being anti-symmetric matrices and $z$ containing the state:
\begin{equation}
 M = \begin{pmatrix}
  0 & -1 & 0 & 0 \\
  1 & 0 & 0 & 0 \\
  0 & 0 & 0 & 0 \\
  0 & 0 & 0 & 0
 \end{pmatrix}\,, \qquad K = \begin{pmatrix}
  0 & 0 & -1 & 0 \\
  0 & 0 & 0 & -1 \\
  1 & 0 & 0 & 0 \\
  0 & 1 & 0 & 0
 \end{pmatrix}\,, \qquad z = \begin{pmatrix}
  a \\
  b \\
  v \\
  w
  \end{pmatrix}.
\end{equation}
This formulation emphasizes the multi-symplectic nature of the system.
By introducing a spatial and timely discretization $x_n, n=1,2,\dots$ and $t_m, m=1,2,\dots$ with grid size $\Delta x$ and time step $\Delta t$ the discretized state is defined $z_n^m := z( x_n , t_m )$.
The multi-symplectic Euler-Box scheme for this discrete state variable is then given as:
\begin{equation} \label{eqn:euler_box}
 M_+ \delta^+ z_n^m + M_-\delta^- z_n^m + K_+\delta_+ z_n^m + K_- \delta_- z_n^m = \nabla_z S(z_n^m),
\end{equation} 
with $\delta^{+-}$, $\delta_{+-}$ are the forward/backward difference operators acting on time and space, e.g.\ $\delta^+ z_n^m = (z_n^{m+1} - z_n^m)/\Delta t$ and $\delta_- z_n^m = (z_{n}^m - z_{n-1}^m)/\Delta x$.
$M_+$, $M_-$ and $K_+$, $K_-$ are splittings of the symplectic structure matrices $M = M_+ + M_-$ and $K = K_+ + K_-$, where the conservation of symplectivity demands $M_+^T = -M_-$ and $K_+^T = -K_-$.
This matrix splitting is not unique. 
However, in~\cite{Wang_Li_Song_08} it was found that all possible splittings lead to one of two fundamental schemes.
The first one is given by taking $M_+$ and $K_+$ as upper triangular matrices:
\begin{equation}
 M_+ = \begin{pmatrix}
  0 & -1 & 0 & 0 \\
  0 & 0 & 0 & 0 \\
  0 & 0 & 0 & 0 \\
  0 & 0 & 0 & 0
 \end{pmatrix}\,, \qquad K_+ = \begin{pmatrix}
  0 & 0 & -1 & 0 \\
  0 & 0 & 0 & -1 \\
  0 & 0 & 0 & 0 \\
  0 & 0 & 0 & 0
 \end{pmatrix}.
\end{equation} 
Submitting this into \eqref{eqn:euler_box} we obtain, after substituting $v_n^m$ and $w_n^m$, the following scheme:
\begin{equation}
\begin{aligned}
 b_n^{m+1} &= b_n^m - \frac{\Delta t}{\Delta x^2} (a_{n+1}^{m} - 2a_n^m + a_{n-1}^m ) - \Delta t(V_n+2) a_n^{m} \\ & \qquad - \Delta t \beta ( (a_n^m)^2 + (b_n^m)^2 ) a_n^m \\
 a_n^{m+1} &= a_n^m + \frac{\Delta t}{\Delta x^2}( b_{n+1}^{m+1} - 2 b_n^{m+1} + b_{n-1}^{m+1}) + \Delta t(V_n+2)a_n^{m+1} \\ & \qquad +\Delta t\beta( (a_n^{m+1})^2 + (b_n^{m+1})^2 ) b_n^{m+1}.
\end{aligned}
\end{equation} 
This is a first order, multi-symplectic scheme~\cite{Wang_Li_Song_08,Hong_Hongyu_Sun_06}.
The scheme is locally implicit as the second equation for $a_n^{m+1}$ is given implicitly and requires to solve a quadratic equation.
However, no implicit dependence on neighboring lattice sites is included and hence the scheme does not involve solving a set of nonlinear equations which makes it \emph{locally} implicit.
From here on, we will set $\Delta x = 1$ which makes this method solving the DNLS~\eqref{eqn:DANSE}.

The second fundamental splitting is obtained by exchanging $M_+ \leftrightarrow M_-$ while leaving $K_+$ and $K_-$ as is.
This gives another scheme similar to the one above but with a locally implicit equation for $b_n^{m+1}$ instead of $a_n^{m+1}$.
It can be shown that the two schemes are adjoint to each other with respect to time reversal~\cite{Wang_Li_Song_08}.
The adjoint method of a one-step method $z^{m+1} = \Phi_{\Delta t} z^{m}$ is defined as the inverse original map with reversed time $-\Delta t$, that is $\Phi^*_{\Delta t} := \Phi^{-1}_{-\Delta t}$ and thus can be obtained by solving the following equations for $z^{m+1}$: $ z^m = \Phi_{-\Delta t}( z^{m+1} )$.
Given the two adjoint schemes above by composition theory a new second-order, self-adjoint method can be created by constructing the mapping $\Phi^*_{\Dt/2}\Phi_{\Dt/2}$~\cite{Leimkuhler_Reich_05}, which is given by the following equations:
\begin{equation}
\begin{aligned}
  a_n^{m+\frac12} &= a_n^m - \frac\Dt2 ( b_{n+1}^{m} - 2 b_n^{m} + b_{n-1}^{m} ) - \Dt (V_n+2) b_n^{m} \\
   & \qquad - \Delta t \beta ( (a_n^m)^2 + (b_n^m)^2 ) b_n^m \\
  b_n^{m+\frac12} &= b_n^m - \frac\Dt2 ( a_{n+1}^{m+\frac12} - 2 a_n^{m+\frac12} + a_{n-1}^{m+\frac12} ) - \Dt (V_n+2) a_n^{m+\frac12} \\
   & \qquad - \Delta t \beta ( (a_n^{m+\frac12})^2 + (b_n^{m+\frac12})^2 ) a_n^{m+\frac12} \\
  b_n^{m+1} &= 2b_n^{m+\frac12} - b_n^m \\
  a_n^{m+1} &= a_n^{m+\frac12} - \frac\Dt2 ( b_{n+1}^{m+1} - 2 b_n^{m+1} + b_{n-1}^{m+1} ) - \Dt (V_n+2) b_n^{m+1} \\
   & \qquad - \Delta t \beta ( (a_n^{m+1})^2 + (b_n^{m+1})^2 ) b_n^{m+1}.
\end{aligned}
\end{equation} 
This scheme consists of two explicit (first and third line) and two locally implicit steps (second and fourth line) and defines the multi-symplectic Euler-Box (EB) scheme.
The behavior of the energy error for this scheme is plotted in Figure~\ref{fig:pq} (black stars).
As expected, it is second order accurate, and the error lies between the PQ$_\text{SBAB1}$ and PQ$_\text{SBAB2}$ schemes.
However, the Euler-Box method requires to evaluate two root functions for each lattice site at each step, which is computationally quite demanding.
So again, only a detailed performance study can reveal if this method presents some advantage over the other schemes introduced above.
It should be mentioned that this scheme is still a particularly interesting approach for solving the DNLS, as it can be easily generalized to two- or even higher-dimensional lattices.
A clear advantage over the other methods described above, for which a generalization to higher dimensions is not straight forward.

\subsection{High-Order Schemes}
All the methods above were accurate to at most second order.
Of course, one often requires higher order methods, especially when a good accuracy is wanted.
Fortunately, there exists a generic way of constructing higher order methods from symmetric, second order schemes such as the SBAB1, SBAB2 or the Euler-Box algorithm~\cite{Leimkuhler_Reich_05}.
Many such schemes have been developed, for example by Yoshida~\cite{Yoshida_90}, but here we restrict ourselves to presenting two higher-order methods, one of order four and one of order six.
Assume we have a symmetric second order scheme defined by the mapping $\psi' = \Phi(\Dt) \psi$ with the stepsize~$\Dt$ as parameter.
In this context, a scheme is called symmetric if it is self-adjoint ${\Phi(\Dt) = \Phi^*(\Dt)}$, where the adjoint scheme is the inverse mapping with negative stepsize: ${\Phi^*(\Dt) = \Phi^{-1}(-\Dt)}$.
For the SBAB1/2 schemes it can be easily seen that they are indeed symmetric, but this also holds for the Euler-Box scheme, by construction, as explained in section~\ref{sec:euler_box} and in~\cite{Wang_Li_Song_08}.

\begin{figure}[t]
  \begin{subfigure}{0.5\textwidth}
    \includegraphics[width=0.95\textwidth]{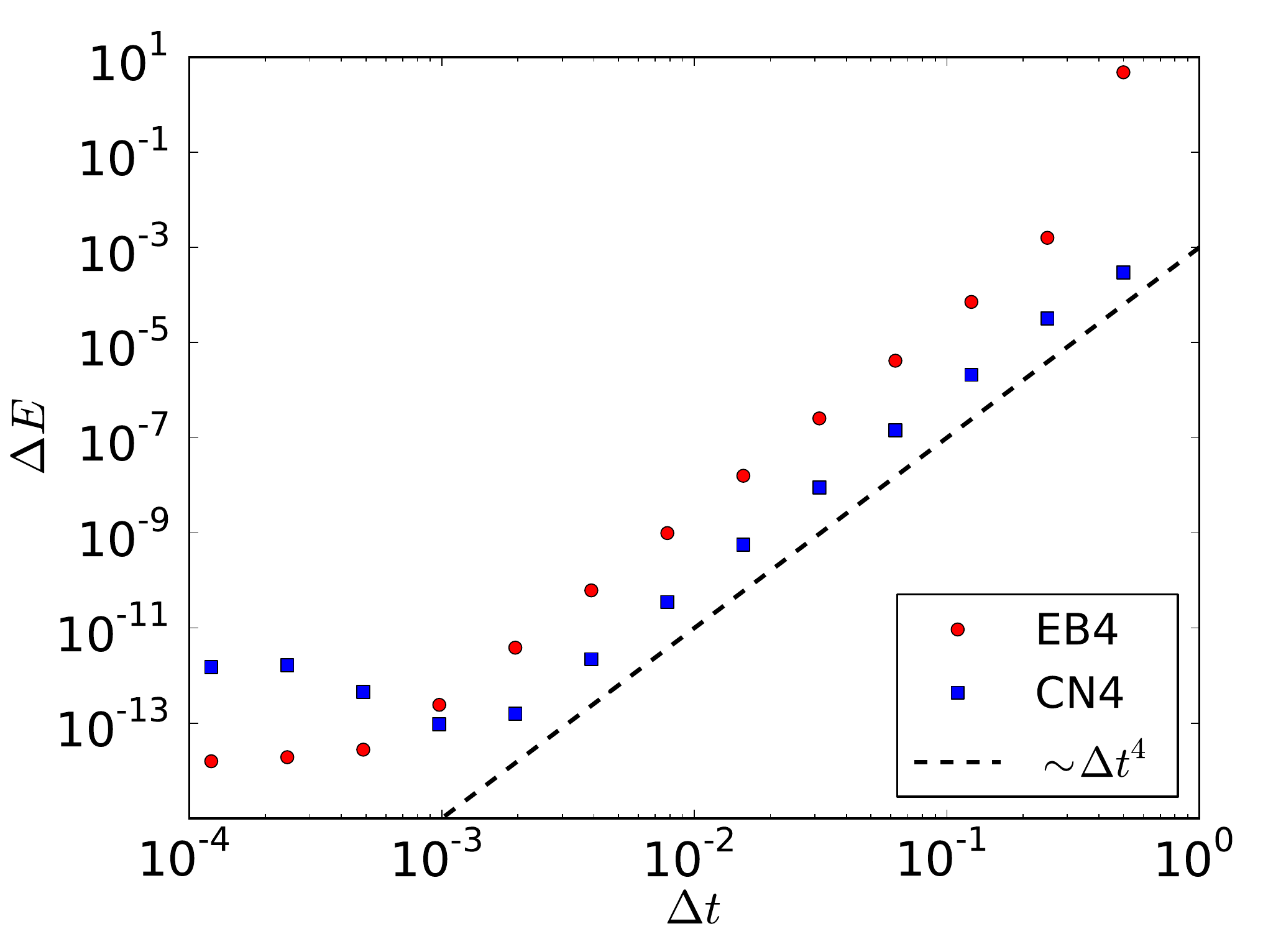}
    \caption{Fourth order Yoshida composition.}
  \end{subfigure}
  \begin{subfigure}{0.5\textwidth}
    \includegraphics[width=0.95\textwidth]{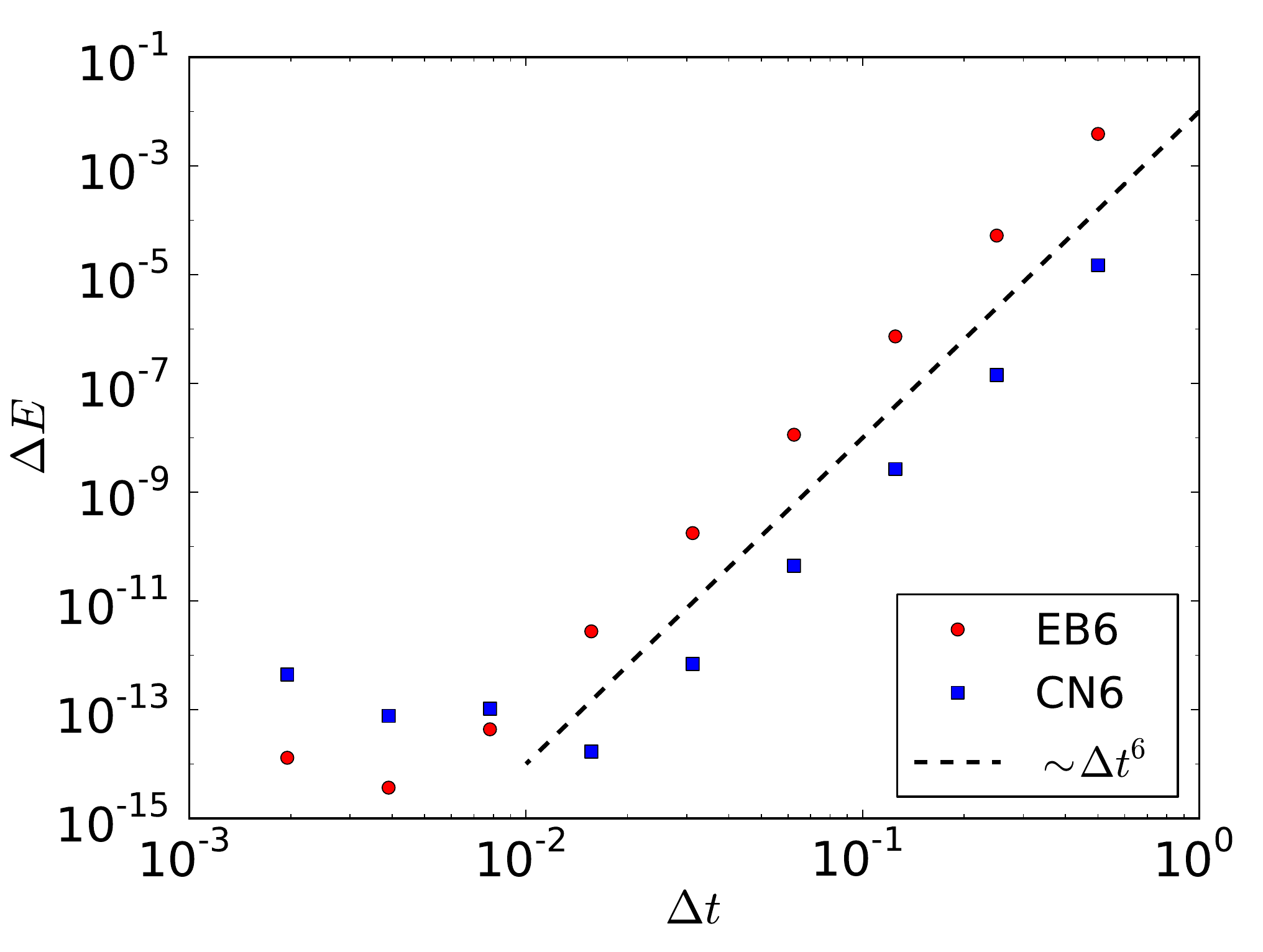}
    \caption{Sixth order Yoshida composition.}
  \end{subfigure}
  \caption{ Energy error of the fourth (a) and sixth (b) order composition methods based on the SBAB2 Crank-Nicolson (CN4/CN6) and the Euler-Box (EB4/EB6) scheme. The numerical setup is the same as for the results above.}
 \label{fig:yo}
\end{figure}

Following~\cite{Yoshida_90}, we construct a fourth order scheme $\Phi^4(\Dt)$ as follows:
\begin{equation} \label{eqn:yoshida_4}
 \Phi^4(\Dt) = \Phi(\alpha_1 \Dt) \Phi(\alpha_0 \Dt) \Phi(\alpha_1 \Dt),
\end{equation}
with $\alpha_0=-2^{1/3}/(2-2^{1/3})$ and $\alpha_1=1/(2-2^{1/3})$~\cite{Yoshida_90}.
Similarly, a sixth order scheme can be constructed:
\begin{equation} \label{eqn:yoshida_6}
\begin{aligned}
 \Phi^6(\Dt) =\; &\Phi(\mu_3 \Dt) \Phi(\mu_2 \Dt) \Phi(\mu_1 \Dt)\Phi(\mu_0 \Dt) \Phi(\mu_1 \Dt) \Phi(\mu_2 \Dt) \Phi(\mu_3 \Dt).
\end{aligned}
\end{equation}
The parameters $\mu_{0\dots3}$ are not defined uniquely in this case.
Here we will use the set called ``solution A'' in~\cite{Yoshida_90}:
\begin{align*}
 \mu_1 &=  -1.17767998417887 &
 \mu_2 &=  0.235573213359357 \\
 \mu_3 &=  0.784513610477560 &
 \mu_0 &= 1 - \mu_1 - \mu_2 - \mu_3.
\end{align*}
Note, that these methods of constructing higher order schemes based on the PQ$_\text{ABC}$ scheme have recently been discussed by Skokos et al.\ in~\cite{Skokos_etal_13}, where they also used the DNLS with random potential as benchmarking example.
In Figure~\ref{fig:yo} we examplarily show the energy error for the schemes~$\Phi^4$ (a) and~$\Phi^6$ (b) based on the symplectic CN$_\text{SBAB2}$ and the Euler Box schemes.
The $\order(\Dt^4)$ and $\order(\Dt^6)$ behavior is clearly visible, again with a smaller error for the CN based schemes.

\section{Performance} \label{sec:performance}

After having introduced and discussed several numerical methods to solve the Discrete Nonlinear Schr\"odinger equation, we will now study the efficiency of the different approaches.
We will first consider only the second order schemes, as the most efficient of these can then also be used to construct the most efficient higher order scheme.
For all of the methods described above the computational effort increases linearly with system size\footnote{The FT method has an $N\log N$ behavior, so it might become more disadvantageous for very large systems}, hence no study on different system sizes is required to compare the different methods.
Note especially that although the Crank-Nicolson (CN) scheme involves solving a linear system of $N$ equations, which formally is of complexity $\order(N^2)$, one can make use of the band structure of the coupling term~$L_{A_\text{CN}}$ with only three non-zero diagonals and thus implement this algorithm with complexity $\order(N)$ as well.

\begin{figure}[t]
  \begin{subfigure}{0.5\textwidth}
    \includegraphics[width=0.95\textwidth]{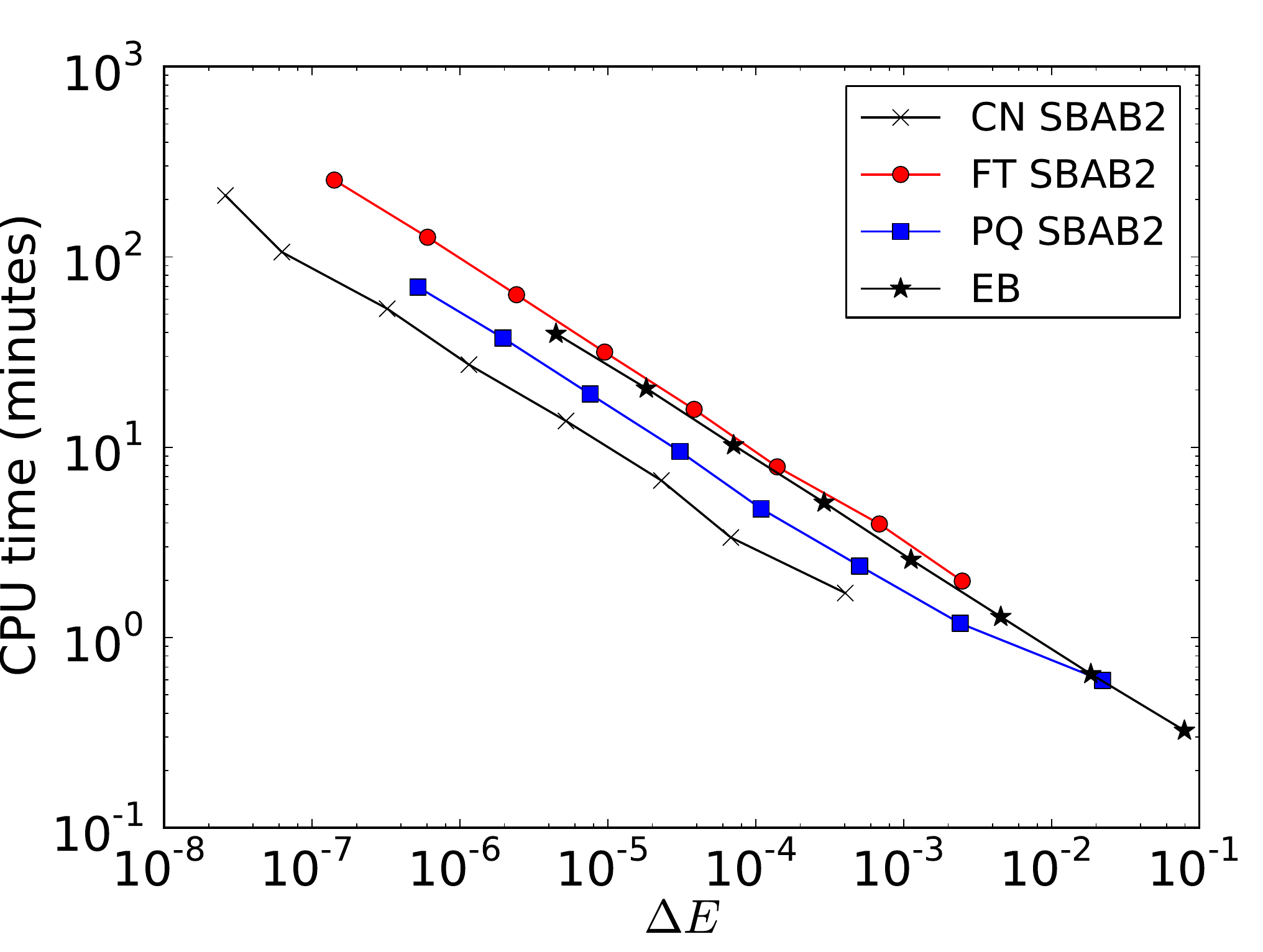} \hfill
    \caption{Performance of second order methods.} \label{fig:perf_2}
  \end{subfigure}
  \begin{subfigure}{0.5\textwidth}
    \includegraphics[width=0.95\textwidth]{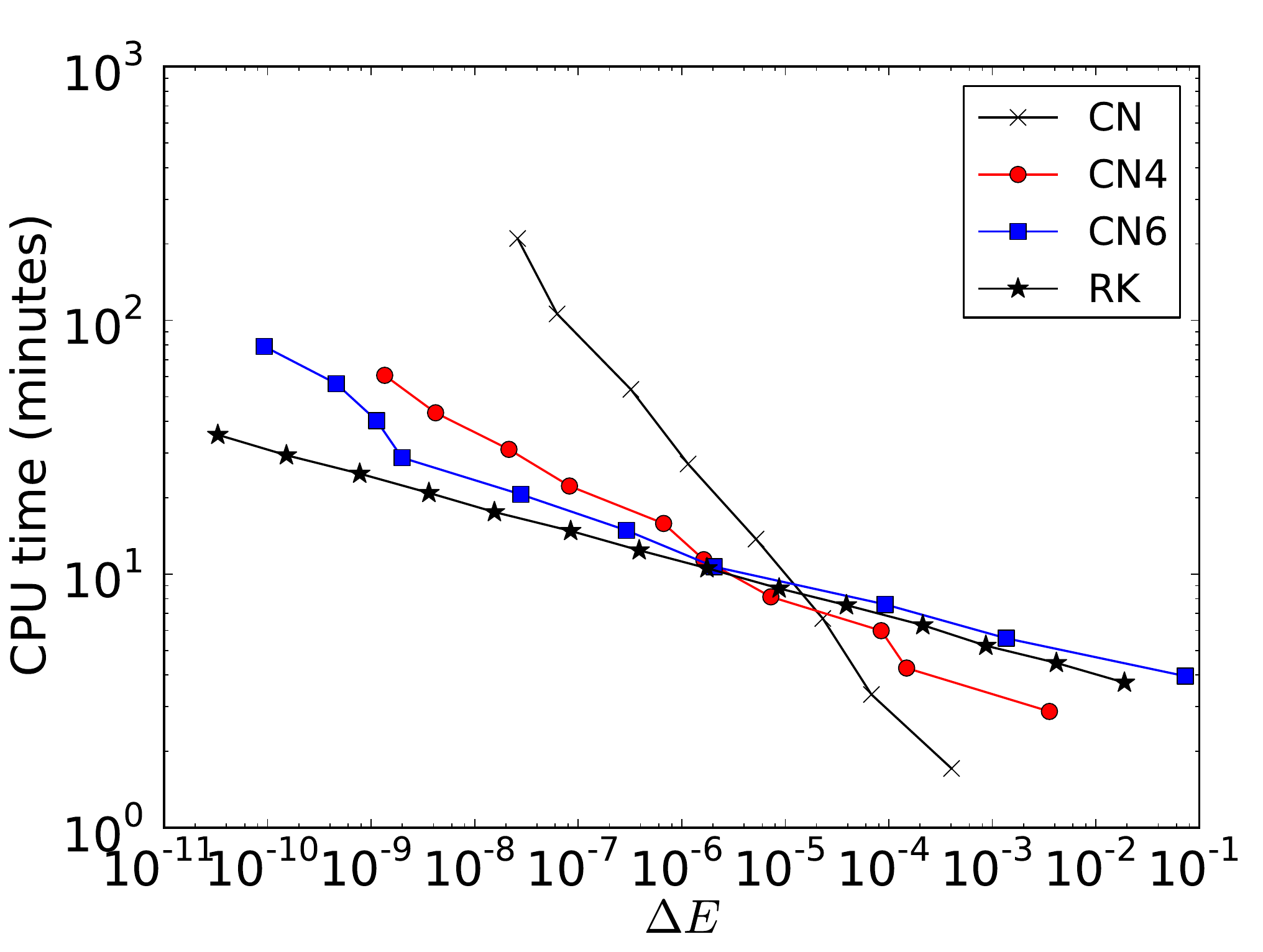} \hfill
    \caption{Performance of higher order methods.} \label{fig:perf_ho}
  \end{subfigure}
  \caption{Performance of the different second order schemes (a) and higher order schemes based on CN$_\text{SBAB2}$ (b). These graph show the CPU time to integrate a system with $N=1024$ lattice sites up to $T=10^5$ with a given averaged energy error~$\Delta E$ on an Intel Core i7, 2.93GHz machine. In (b), additionally the results for a non-symplectic 8-th order Runge-Kutta scheme (RK) are plotted.}
  \label{fig:performance}
\end{figure}

We analyze the performance of the different methods by computing a trajectory for a lattice with $N=1024$ sites and a random potential $V_n\in[-2,2]$ and nonlinear strength~$\beta=1$.
This is the typical setup for studying spreading in the DANSE model~\cite{Pikovsky_Shepelyansky_08,Flach_Krimer_Skokos_09,Mulansky_Pikovsky_10}.
We again start the time evolution from an initial Gaussian with width~$\sigma=10$ and total norm $\norm=1$.
We compute a trajectory up to a time $T=10^5$ with decreasing stepsizes~$\Dt$.
As before, we use the mean squared energy error $\Delta E$ as given in~\eqref{eqn:dE} as quantification of the accuracy, but here the energy is only computed every 10 timesteps, so the major computational effort lies in the time evolution.
The results for the different second order schemes are shown in Figure~\ref{fig:perf_2}, where the CPU time is plotted in dependence of the mean energy error~$\Delta E$.
The FT$_\text{SBAB2}$ and the EB exhibit a similar performance, while the PQ$_\text{SBAB2}$ scheme introduced by Skokos et al.~\cite{Skokos_etal_09} represents a clear improvement.
However, one gets the best performance when using the CN$_\text{SBAB2}$ scheme.
It requires about the same computational effort as the FT$_\text{SBAB2}$ scheme, but its different splitting gives a much smaller energy error, as already seen from comparing Figures~\ref{fig:ft} and \ref{fig:cn}.
We also checked the performance of other splittings (SBAB$_1$ and ABC), but the SBAB$_2$ versions were always superior.

In Figure~\ref{fig:perf_ho} we compare the performance of higher order schemes based on the most efficient second order scheme CN$_\text{SBAB2}$.
Namely, we show results for the Yoshida 4 and Yoshida 6 compositions introduced in~\eqref{eqn:yoshida_4} and~\eqref{eqn:yoshida_6}.
For comparison, we also plot the second order results for CN$_\text{SBAB2}$.
Interestingly, the higher order schemes are only more efficient when a high accuracy is required.
If an accuracy $\Delta E>10^{-5}$ is sufficient, the second order scheme CN$_\text{SBAB2}$ provides the best performance.
If one requires smaller errors, a higher order scheme should be chosen.
In Figure~\ref{fig:perf_ho} also the performance of a non-symplectic 8-th order Runge-Kutta scheme with stepsize control~\cite{Hairer_Norsett_Wanner_93} is presented, as implemented in the Boost.Odeint library~\cite{Ahnert_Mulansky_11}.
We found that for the total integration time chosen here, $T=10^5$, this non-symplectic scheme is competitive and even the most efficient for very high accuracy.
However, one has to keep in mind that this scheme is non-symplectic, hence the energy error $\Delta E$ will increase linearly in time.
Thus, for very long time scales $T>10^5$, the symplectic schemes will outperform the controlled Runge-Kutta scheme.
For the DANSE model that was used here examplarily, the integration times go up to $T=10^9$~\cite{Mulansky_dipl_09}, where the higher order symplectic schemes will surely beat the RK method.

\section{Conclusions} \label{sec:conclusions}
We introduced and described several numerical schemes to compute approximate trajectories for the Discrete Nonlinear Schr\"odinger equation~\eqref{eqn:DANSE}.
The DNLS is a very important model with many applications, hence identifying the most efficient numerical scheme is of great interest.
For our numerical performance test we relied on the DANSE setup that is used extensively in the past to study the interplay between disorder and nonlinearity.
Of the mostly used second order splitting schemes, the CN$_\text{SBAB2}$ method was found to show the best performance in terms of the least CPU time for a given energy error~$\Delta E$.
Hence, we conclude that this scheme is superior to the others and should be the first choice for numerically treating DNLS models when a moderate precision is required.
A new and particularly interesting approach is the Euler-Box scheme as it can also be used for higher dimensional systems.
But the implicit treatment of the nonlinearity in this case strongly restricts the possibility to treat different local nonlinear potentials (e.g.\ higher powers).
The FT, CN and PQ splitting schemes, on the other hand, can be used for any local nonlinear potential in the DNLS and are thus more flexible in this regard.
Note, that all presented schemes here are suitable to additionally integrate the set of linearized equations to obtain Lyapunov exponents.
For the CN scheme this has been already done in~\cite{Mulansky_dipl_09}.

Finally, we showed that when a high accuracy is required, higher order schemes provide better performance than the standard second order schemes.
At an integration time of $T\approx10^5$, even non-symplectic schemes are competitive.
But this changes with increasing $T$.
We note that performance results are always to be taken with care as they might vary greatly for different compilers, operating systems or hardware.
However, we believe that our results are a helpful guide for which simulation code to choose when dealing with DNLS models.

\section*{Acknowledgements}
I thank A.~Pikovsky and B.~Mulansky for fruitful discussions.
I also thank the CCT at Louisiana State University for hospitality and financial support as well as the DAAD for financial support under the project number 50015188.
Finally, financial support under the Project HPC-EUROPA2 (Project number 228398), with the support of the European Community - under the FP7 ``Research Infrastructure'' Programme is acknowledged.

\bibliographystyle{siam}
\bibliography{sdm}

\end{document}